\documentclass[pre, aps, showpacs, superscriptaddress, twocolumn, longbibliography]{revtex4-2}

\usepackage{hyperref}
\hypersetup{
     colorlinks=true,
     linkcolor=magenta,
     filecolor=blue,
     citecolor=blue,      
     urlcolor=cyan,
     }
\usepackage{color}
\usepackage[usenames,dvipsnames]{xcolor}
\usepackage{amsmath,amsthm,amssymb}
\usepackage{graphicx}
\usepackage{float}
\usepackage{epsfig}
\usepackage{bm}
\usepackage{mathrsfs}
\usepackage{multirow}
\usepackage[all]{xy}
\usepackage{pbox}
\usepackage{verbatim}
\usepackage{braket}
\usepackage{mathtools}
\usepackage{bm}
\usepackage{tikz}
\usepackage{xcolor}
 
\usepackage{mathtools}

\usepackage{mathtools}
\usepackage{tabstackengine}
\usepackage{enumerate}   
\stackMath
\DeclareMathOperator{\Tr}{Tr}

\newcommand{\er}[1]{Eq.~\eqref{#1}}

\newcommand{\era}[2]{Eqs.~(\ref{#1}) and (\ref{#2})}

\newcommand{\beq}{\begin{equation}}
\newcommand{\eeq}{\end{equation}}

\newcommand{\N}{\mathcal N}
\newcommand{\D}{\mathcal D}
\renewcommand{\O}{\mathcal O}
\newcommand{\sst}{\ket{\text{ss}}}
\newcommand{\rra}[1]{( #1 |}
\newcommand{\ret}[1]{| #1 )}

\newcommand{\W}{{\mathbb W}}
\renewcommand{\H}{{\mathbb H}}

\begin{document}  

\title{Slow dynamics and large deviations in classical stochastic Fredkin chains}

\author{Luke Causer}
\affiliation{School of Physics and Astronomy, University of Nottingham, Nottingham, NG7 2RD, UK}
\affiliation{Centre for the Mathematics and Theoretical Physics of Quantum Non-Equilibrium Systems,
University of Nottingham, Nottingham, NG7 2RD, UK}
\author{Juan P. Garrahan}
\affiliation{School of Physics and Astronomy, University of Nottingham, Nottingham, NG7 2RD, UK}
\affiliation{Centre for the Mathematics and Theoretical Physics of Quantum Non-Equilibrium Systems,
University of Nottingham, Nottingham, NG7 2RD, UK}
\author{Austen Lamacraft}
\affiliation{TCM Group, Cavendish Laboratory, University of Cambridge, Cambridge CB3 0HE, UK}

\begin{abstract}
    The Fredkin spin chain serves as an interesting theoretical example of a quantum Hamiltonian whose ground state exhibits a phase transition between three distinct phases, one of which violates the area law.
    Here we consider a classical stochastic version of the Fredkin model, which can be thought of as a simple exclusion process subject to additional kinetic constraints,  
    and study its classical stochastic dynamics. 
    The ground state phase transition of the quantum chain implies an equilibrium phase transition in the stochastic problem, whose properties we quantify in terms of numerical matrix product states (MPS). The stochastic model displays slow dynamics, including power law decaying autocorrelation functions and hierarchical relaxation processes due to exponential localization.
    Like in other kinetically constrained models, the Fredkin chain has a rich structure in its dynamical large deviations - 
    which we compute accurately via numerical MPS -  
    including an active-inactive phase transition, and a hierarchy of trajectory phases connected to particular equilibrium states of the model. We also propose, via its height field representation, a generalization of the Fredkin model to two dimensions in terms of constrained dimer coverings of the honeycomb lattice. 
\end{abstract}

\maketitle

\section{Introduction}
The Fredkin spin chain \cite{Salberger2016, Movassagh2016} is a one-dimensional lattice model with local three-body interactions, whereby hardcore particles can hop to adjacent sites if allowed by constraints involving next to nearest neighbors. This model has been of interest in the quantum many-body community over the last few years for a number of reasons. In its original formulation \cite{Salberger2016, Movassagh2016}, the Fredkin chain can be expressed exactly as an equal superposition of all Dyck paths, i.e.\ random walk (RW) excursions, with appropriate endpoints, with an entanglement entropy which scales logarithmically in system size, thus violating the area law \cite{Movassagh2016, Salberger2016, Sugino2018, DellAnna2019a}.
Furthermore, the model has slow unitary evolution \cite{DellAnna2016, Chen2017, DellAnna2019b, Adhikari2020} due to dynamical ``jamming''. With the addition of particular potential energy terms the model features a ground state phase transition between states of bounded and extensive entanglement entropy \cite{Salberger2017, Udagawa2017}. These interesting properties have brought about further studies into generalized Fredkin models \cite{Adhikari2019}, including versions which present ``quantum scars'' \cite{Langlett2021}.

Dynamical constraints, such as those present in the Fredkin model, are responsible more generally for many interesting phenomena in many-body dynamics.
A striking example of this are the kinetically constrained models (KCMs) of structural glasses \cite{Ritort2003, Garrahan2011} - simple lattice models equipped with local dynamical constraints, leading to slow relaxation and dynamical heterogeneity \cite{Garrahan2002, Chandler2010}. Such models can also be considered as systems under closed unitary \cite{Horssen2015, Prem2017, Pancotti2020, Pozsgay2021} and open dissipative  \cite{Olmos2012, Olmos2014, Rose2020b} quantum dynamics.
A recent example of these is the quantum PXP model  \cite{Fendley2004,Lesanovsky2011}
of Rydberg atoms in optical lattices under blockade conditions, which has been shown to exhibit non-thermal eigenstates (often called quantum scars \cite{Turner2018}).
Another area where dynamical constraints lead to interesting non-equilibrium dynamics is in deterministic cellular automata 
\cite{Prosen2016, Inoue2018, Prosen2017, Buca2019, Friedman2019, Gopalakrishnan2018, Gopalakrishnan2018b, Klobas2019, Klobas2020b, Alba2019, Alba2020, Klobas2020, Iadecola2020, Wilkinson2020, Gombor2021, Wilkinson2021, Klobas2021, Pozsgay2021a} (for a review see \cite{Buca2021}). Recently, cellular automata circuits have been also used to study Fredkin-like systems \cite{Singh2021, Richter2021}, revealing a new ``universality class'' of hydrodynamics. While the connection of the Fredkin quantum spin chain to stochastic dynamics has been previously mentioned \cite{Movassagh2018}, it has not yet been extensively explored (other than briefly in \cite{Brewer2019}). Here we provide such systematic study of both typical dynamics and rare fluctuations.

\begin{figure}[t]
    \centering
    \includegraphics[width=\linewidth]{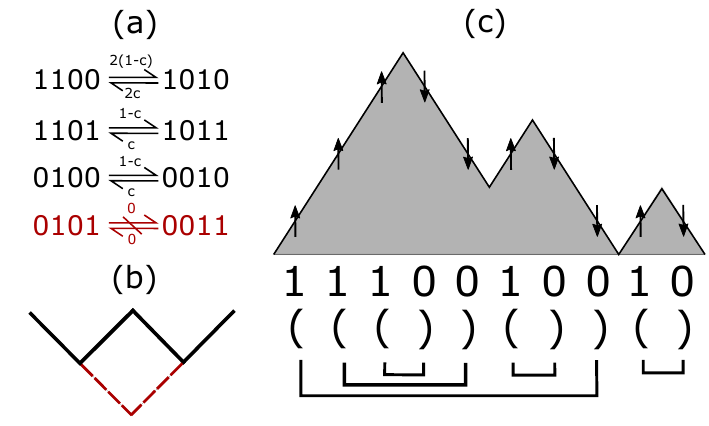}
    \caption{\textbf{Fredkin spin chains.} 
    (a) The local stochastic transition rates for neighboring occupied and unoccupied sites, given by all choices of their neighbors. The fourth transition is not allowed.
    (b) The disallowed configuration change in the height representation. The ``troughs'' ($\cdots0011\cdots$) are locally immobile.
    (c) An example configuration in the chosen symmetry sector. The top shows the RW representation of the height field, which must always satisfy $h > 0$. The middle is the corresponding particle representation. The bottom is in terms of Dyck words, where opening ``('' must always be matched with a closing ``)''.
    }
    \label{fig: representations}
\end{figure}

Classically, the Fredkin model resembles the simple exclusion process (or SEP, for reviews see \cite{Blythe2007, Mallick2015}). Like the SEP, it describes particles hopping stochastically to neighboring empty lattice sites with at most one particle per site. The key difference is the presence of further local kinetic constraints to motion. These, together with specific boundary conditions, specifically that of an open segment with fixed boundaries, restrict the dimensionality of the state space. For example, for a length $N=2M$ chain half filled with $M$ particles the dimensionality is the Catalan number $C_N=\frac{1}{M+1}\binom{2M}{M}$ rather than the binomial coefficient $\binom{2M}{M}$. Although the difference in configurational entropy is not extensive, this constrained state space plays an important role in the dynamics, as we explain below. 

SEPs and KCMs display interesting dynamical properties which can be studied with large deviation (LD) methods (for reviews see \cite{Touchette2009, Garrahan2018, Jack2020, Limmer2021}). A central result in the dynamics of these systems is the existence of phase transitions in the space of trajectories, indicated by singularities in the LD functions that quantify the dynamical fluctuations in the long time limit, both in terms of time-integrated currents \cite{Bodineau2007, Appert-Rolland2008, Bodineau2012, Karevski2017}, or dynamical activities \cite{Merolle2005, Garrahan2007, Lecomte2007, Garrahan2009, Jack2015}. In the case of the Fredkin model, a preliminary study \cite{Brewer2019} indicated that it also displays LD transitions. Here we make this finding concrete by studying LD functions using matrix product states.

The paper is organized as follows. 
In Sec.~\ref{model} we start by defining the model and reviewing its basic properties. We highlight its relationship to Catalan combinatorics and RW excursions \cite{Majumdar2005}. 
In Sec.~\ref{equilibrium} we consider the equilibrium states which follow from the properties of the ground state of the quantum problem \cite{Salberger2016, Salberger2017}. We study the properties of the equilibrium phases in detail by means of numerical DMRG \cite{White1992}. An interesting observation is that there are three distinct equilibrium phases, and a transition between them, despite the fact that this is a one-dimensional system with local dynamical rules. This apparent contradiction with the Landau principle is a consequence of the constrained configuration space of the model. 
In Sec.~\ref{reldyn} we study the relaxation dynamics. As in the case of the quantum model \cite{Adhikari2020}, the stochastic Fredkin spin chain exhibits slow dynamics. We provide evidence for power law decaying autocorrelations, and for a pattern of hierarchical relaxation when quenched from extremal initial states into the different equilibrium phases. 
In Sec.~\ref{dynLDs} we study the large deviations statistics of dynamical observables by means of numerical MPS. As in other constrained models, the phase transitions at the LD level underpin the slow dynamics and fluctuations seen in typical relaxation trajectories.
We reveal the existence of an active-inactive transition, as in other KCMs, but also a hierarchy of trajectory transitions connected to hierarchical relaxation dynamics. In Sec.~\ref{tilings} we speculate on a possible generalization of the Fredkin model to a two-dimensional setting defined in terms of fully packed dimers on the honeycomb lattice (that is, rhombus coverings of the plane). We give our conclusions in Sec.~\ref{conc}.

\section{Model} \label{model}
The Fredkin model is defined in terms of particles hopping on a lattice of $N$ sites with binary occupation, $n_{j} = 0$ (for empty or down) or $1$ (for occupied or up) with $j = 1, \cdots, N$.
The system evolves under stochastic continuous-time Markov dynamics with generator
\begin{align}
    \W = \sum_{i = 2}^{N-2} 
        &
        f_{i} 
        \left\{    
            c 
            \left[
                 \sigma^{+}_{i} \sigma^{-}_{i+1} 
                 - (1-n_{i}) n_{i+1} 
            \right]
        \right.
    \nonumber
    \\ 
        & 
        \phantom{f_{i}}
        \left.
            + (1-c) 
            \left[
                \sigma^{-}_{i} \sigma^{+}_{i+1} 
                - n_{i} (1-n_{i+1}) 
            \right] 
        \right\} ,
    \label{W}
\end{align}
where $\sigma^{\pm}_{i}$ are Pauli creation and annihilation operators on site $i$. The factor in curly brackets in each term is the same as the local generator of the asymmetric SEP (ASEP) \cite{Blythe2007, Mallick2015}, with rates for hops to the left or right given by $c$ and $1-c$, respectively. What distinguishes the Fredkin model from the ASEP is the kinetic constraint
\beq
    f_{i} = n_{i-1}  + (1 - n_{i+2}) ,
    \label{fi}
\eeq
which means that hopping between sites $i$ and $i+1$ is not allowed if $n_{i-1} = 0$ and $n_{i+2} = 1$, see Fig.~\ref{fig: representations}(a)
\footnote{
    Note that the Fredkin model is different from the so-called ``facilitated'' SEP \cite{Gabel2010}, where jumps are only allowed if there is a particle behind the one that is hopping. For example, in  that model the third transition in Fig.~1(a) is not allowed.
}. 
In \er{W} we are considering open boundary conditions on a segment $[1,N]$ with no injection/ejection of particles at the boundaries. The fixed sites at the edges, which are not acted on by the generator, we fix to be $n_{1} = 1$ and $n_{N} = 0$.

Note that at $c = 1/2$, \er{W} is equivalent to the quantum Hamiltonian of the original Fredkin model defined in Ref.~\cite{Salberger2016}, up to a minus sign and boundary terms. For $c \neq 1/2$ the generator \er{W} obeys detailed balance despite the asymmetry in the hopping rates
\footnote{
    This is analogous to what occurs in the ASEP with the same kind of boundaries, see e.g. \cite{De-Gier2006}. 
}, meaning that for all values of $c$ we expect to find an equilibrium stationary state of $\W$. Notice that under a similarity transformation (see below) it becomes equivalent to the ``deformed'' Fredkin model of Ref.~\cite{Salberger2017}.  

\begin{figure*}[t]
    \centering
    \includegraphics[width=\linewidth]{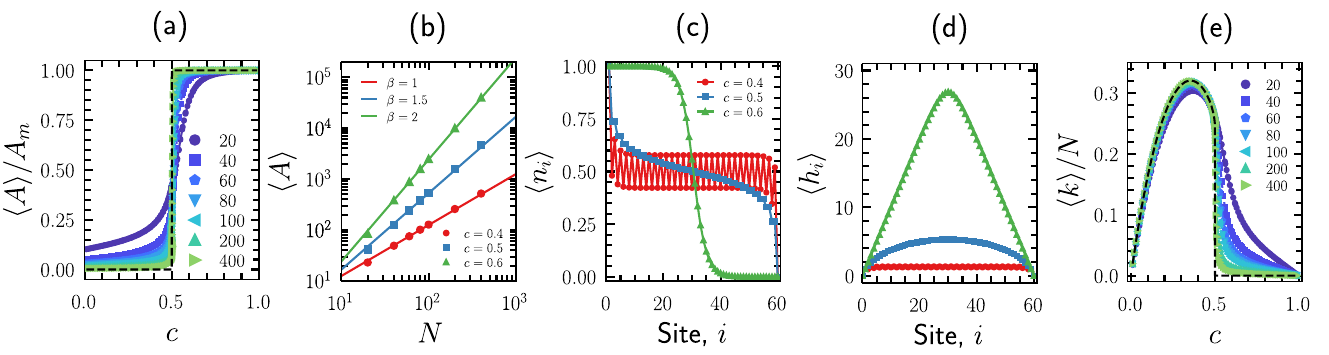}
    \caption{\textbf{Equilibrium properties of the Fredkin model.}
    (a) The average area (scaled by maximum area) $\braket{A} / A_{m}$ as a function of $c$ for various systems sizes $N \in [20, 400]$. The dashed line shows the extrapolated value for $N\to\infty$.
    (b) The average area (symbols) for $c = 0.4$ (red / dark grey), $c = 0.5$ (blue / medium grey) and $c = 0.6$ (green / light grey). The lines show the power laws $\braket{N} \sim N^{-\beta}$ with $\beta = 1,\, 1.5,\, 2$ respectively.
    (c, d) The spin occupation $\braket{n}_{i}$ and height profiles $\braket{h_{i}}$ for each equilibrium phase with a system size $N = 60$.
    (e) The average dynamical activity (per unit time and system size) as a function of $c$ for various systems sizes $N \in [20, 400]$. The dashed line shows the extrapolated value for $N\to\infty$.
    All results are calculated using numerical DMRG.
    }
    \label{fig: equilibrium}
\end{figure*}

The model discussed here has various symmetries.
The most obvious one is the conservation of the total number or particles (or occupied sites): $M = \sum_{i} n_{i}$. This property is shared with the SEP. The constraint \er{fi} gives rise to a further subdivision of each subspace of fixed $M$, which is most easily understood by a representation of the allowed moves in terms of matched brackets \cite{Salberger2016}. In this representation, particles and holes correspond to opening and closing parentheses, and the dynamics respects normal matching rules. Thus the move
\begin{align}
    \cdots 0101\cdots &\longleftrightarrow \cdots 0011 \cdots \\
    \cdots )()(\cdots &\longleftrightarrow \cdots ))(( \cdots 
\end{align}
is forbidden because both sides cannot simultaneously be matched configurations [this forbidden transition is shown in Fig.~\ref{fig: representations}(b)].
Thus a complete specification of a subspace of allowed configurations involves specifying the $M$ pairs of matched brackets, $a$ unmatched opening brackets (particles) and $b$ unmatched closing brackets (holes) for a total $N=2M+a+b$

For concreteness, here we will focus on the case of half-filling by fully matched particles and holes i.e. $M = N / 2$ $a=b=0$. In this case the accumulated number of particles starting from the left is never smaller than the accumulated number of holes (that is, the sector that is dynamically connected to having all particles to the left and all holes to the right, see below), cf.\ \cite{Salberger2016}. We call this sector $\D$.

It is convenient to represent a configuration $x = n_{1:N}$ also in terms of a {\em height field} defined as 
\beq
    h_{i}(x) = \sum_{j=1}^{i} Z_j = h_{i-1}(x) + Z_i
\eeq
with boundary condition $h_{0} = 0$, and where $Z_i = 2 n_i - 1$. For all configurations $x \in \D$ we have $h_{i}(x) \geq 0$ for all $i$. If we think of the space direction as ``time'' and a particle (hole) representing a step up (down), then $\D$ is the space of all paths that correspond to {\em random walk excursions} \cite{Majumdar2005}, that is, random paths that return to the origin while never crossing the horizontal axis. (In contrast, for the SEP in the height representation at half-filling, the space of  dynamically connected configurations is that of random walk bridges, which are also constrained to return to the origin but can cross the horizontal axis.). 
Excursions are also known as Dyck paths.
An example configuration is shown in Fig.~\ref{fig: representations}(c) with each of the representations.

\section{Equilibrium statics} \label{equilibrium}

To determine the equilibrium properties of the model we need to find the state $\sst$ annihilated by \er{W}.
Let us consider as an observable the area under the height profile of a configuration $x$,
\beq
    A(x) = \sum_{i=1}^{N} h_{i}(x) = \sum_{i=1}^{N} (N+1-i) \, Z_i
    \label{area_config} .
\eeq
It is then easy to see that the the stationary state to the dynamics \er{W} is given by \cite{Salberger2017} 
\beq
    \sst = \N_c \sum_{x\in\D} \left(\frac{c}{1-c}\right)^{\frac{1}{2}A(x)} \ket{x} ,
    \label{ss}
\eeq
with $\N_c$ a $c$-dependent normalization constant to make $\langle - \sst = 1$,  where $\bra{-}=\sum_{x\in\D} \bra{x}$ is called the {\em flat state}.

The connection to RW excursions means that this probability distribution is related to the Airy function \cite{Majumdar2005, Agranov2020}.
The properties of the stationary state at arbitrary $c$ can also be understood from the properties of the ground state of the corresponding quantum model \cite{Salberger2016, Salberger2017}. That is, if we perform a similarity transformation of \er{W} (cf.\ the ASEP with the same boundary conditions, e.g.\ \cite{De-Gier2006})
\beq
    \H = -\mathbb{P}^{-1/2} \, \mathbb{W} \, \mathbb{P}^{1/2} ,
    \label{sim}
\eeq
where $\mathbb{P}^{1/2}$ is the diagonal matrix of the square root of configuration probabilities,
\beq
    \braket{x|\mathbb{P}^{1/2}|x} = 
        \N_c^{1/2} 
        \left(
            \frac{c}{1-c}
        \right)^{\frac{1}{4}A(x)} .
    \label{Phalf}
\eeq
we get the Hamiltonian
\begin{align}
    \H &= 
        - \sum_{i=2}^{N-2} 
            f_{i} 
            \left[
                \sqrt{c(1-c)} 
                \left(
                    \sigma^{+}_{i} \sigma^{-}_{i+1} + \sigma^{-}_{i} \sigma^{+}_{i+1}
                \right)
            \right.
    \label{H}
    \\
            & 
            \left.    
                \phantom{\sqrt{c(1-c)}}
                - c (1 - n_{i}) n_{i+1}
                - (1 - c) n_{i} (1 - n_{i+1}) 
            \right] ,
    \nonumber
\end{align}
whose ground state is $\ket{\psi} = \mathbb{P}^{1/2} \sum_{x\in\D}\ket{x}$.
The transformation to a Hermitian form shows that, despite the asymmetric hopping when $c \neq 1/2$, the Fredkin model obeys detailed balance and consequently the stationary state \er{ss} is an equilibrium one.

The properties of the ground state of \er{H} are well understood from previous  studies \cite{Udagawa2017, Adhikari2019}. Here we restate them from the point of view of the equilibrium state of the stochastic model, using matrix product states (MPS, see reviews e.g. Refs.~\cite{Verstraete2008, Schollwoeck2011, Cirac2021}).

\subsection{Exact equilibrium MPS at $c=1/2$}

From the connection to RW excursions at $c=1/2$ the equilibrium state $\sst$ can be written exactly as an MPS 
\begin{align}
    \sst &= \sum_{\{ n_{1:N} \}} 
        \rra{i} 
            B_{n_1}^{(1)} \cdots B_{n_N}^{(N)}
        \ret{f}
        \ket{n_{0:1}}
        \label{MPSeq}
\end{align}
where $B_n^{(j)}$ are site dependent tensors, and $\rra{i}$ and $\ret{f}$ appropriate boundary vectors in the auxiliary (or bond) space of the MPS (we use rounded brackets to distinguish them from vectors in configuration space). 

Consider first the slightly simpler problem of the symmetric SEP (SSEP), whose generator is given by an operator like \er{W} but without a constraint, $f_i=1$. If we consider the same boundary conditions as for the Fredkin model, but with extra terms in \er{W} that allow particle hops between sites $j=1,2$ and $N-1,N$ (no longer prevented in the absence of a constraint), then the SSEP configurations at half-filling are those of RW bridges. If the height field $h_j$ describes the position of the RW after step $j$, the exact transition probabilities at step $j$ for generating bridges of $N$ steps are:
\begin{align}
    T_j^{\rm br}(h \to h \pm 1) = 
        \frac{1}{2} 
        \left( 
            1 \mp \frac{h}{N+1-j}   
        \right)
    \label{Tbr}
\end{align}
for $|h| \leq N+1-j$, or zero otherwise. (These are obtained from the naive symmetric RW transition probabilities via a Doob transform, see e.g.\ \cite{Chetrite2015}.)
The equilibrium MPS for the SSEP is then given by the $(2N+1) \times (2N+1)$ matrices 
\begin{align}
    B_0^{(j), {\rm SSEP}} &= 
        \sum_{h = -N}^N 
            \ret{h} \rra{h-1} \, T_j^{\rm br}(h \to h - 1)
    \label{B0SSEP}
    \\
    B_1^{(j), {\rm SSEP}} &= 
        \sum_{h = -N}^N 
            \ret{h} \rra{h+1} \, T_j^{\rm br}(h \to h + 1)
    \label{B1SSEP}
\end{align}
with boundaries $\rra{i} = \rra{0}$ and $\ret{f} = \ret{0}$.
It is easy to see that the matrices above satisfy $B_0^{(j), {\rm SSEP}} B_1^{(j+1), {\rm SSEP}} - B_1^{(j), {\rm SSEP}} B_0^{(j+1), {\rm SSEP}} = 0$ for all $j$, which means that the MPS \er{MPSeq} with tensors \era{B0SSEP}{B1SSEP} is annihilated by the SSEP generator. 

The construction for the equilibrium state of the stochastic Fredkin chain at $c=1/2$ is similar, but the relevant paths are RW excursions. In this case the ``Doob'' transition probabilities that guarantee an excursion are (cf.\ e.g.\ \cite{Chetrite2015})
\begin{align}
    T_j^{\rm ex}(h \to h \pm 1) = 
        \left\{
            \begin{array}{l}
                \frac{1}{2} 
                \left( 
                    1 + \frac{1}{h+1}   
                \right)
                \left( 
                    1 - \frac{h}{N+1-j}   
                \right) \\
                \\
                \frac{1}{2} 
                \left( 
                    1 - \frac{1}{h+1}   
                \right)
                \left( 
                    1 + \frac{h+2}{N+1-j}   
                \right) \\
            \end{array}
        \right.
    \label{Tex}
\end{align}
for $0 \leq h \leq N+1-j$, or zero otherwise. The corresponding matrices have now bond dimension $N+1$ and read
\begin{align}
    B_0^{(j)} &= 
        \sum_{h = 0}^N 
            \ret{h} \rra{h-1} \, T_j^{\rm ex}(h \to h - 1)
    \label{B0Fredkin}
    \\
    B_1^{(j)} &= 
        \sum_{h = 0}^N 
            \ret{h} \rra{h+1} \, T_j^{\rm ex}(h \to h + 1)
    \label{B1Fredkin}
\end{align}
with the same boundary vectors $\rra{i} = \rra{0}$ and $\ret{f} = \ret{0}$. The relevant relations in this case are 
$B_1^{(j-1)} B_0^{(j)} B_1^{(j+1)} - B_1^{(j-1)} B_1^{(j)} B_0^{(j+1)} = 0$ and $B_0^{(j-1)} B_1^{(j)} B_0^{(j+1)} - B_1^{(j-1)} B_0^{(j)} B_0^{(j+1)} = 0$ for all $j$. Given these, one can show that the MPS \er{MPSeq} with tensors \era{B0Fredkin}{B1Fredkin} is annihilated by the Fredkin generator \er{W}. In fact, the MPS is annihilated by every local term in the spatial sum that defines \er{W}, so that $\W$ can be said to be a {\em parent generator} (cf.\ parent Hamiltonian \cite{Cirac2021}) of the MPS \er{MPSeq}. 

Note that from the definition of the tensors $B_n^{(j)}$ above in terms of transition probabilities, the MPS is in ``right canonical'' form, and \er{MPSeq} therefore satisfies $\langle - \sst = 1$. Away from $c=1/2$ we can also write \er{ss} as an MPS if we reweigh the coefficients in \era{B0Fredkin}{B1Fredkin} as
\begin{align}
    T_j^{\rm ex}(h \to h \pm 1) \to 
        \left( 
            \frac{c}{1-c} 
        \right)^{-\frac{1}{2}(h \pm 1)}
        T_j^{\rm ex}(h \to h \pm 1)
        \, .
        \nonumber
\end{align}
These reweighed coefficients are not transition probabilities in the height (they do not add up to one), meaning that the resulting MPS is not in canonical form. Finding the normalization $\N_c$ in this case is non-trivial.

\subsection{Equilibrium phase diagram from numerical MPS}

To overcome the difficulty above, in order to study the equilibrium properties for all $c$ we resort to numerical MPS approximations. This we implement with the ITensor library \cite{itensor}, and make use of the density matrix renormalization group (DMRG) \cite{White1992, Schollwoeck2005, McCulloch2007} to find the leading eigenvector of \er{H}. We employ an adaptive bond dimension, which is at most $D = 2000$ with a truncation cutoff error $\epsilon = 10^{-12}$.
Furthermore, we exploit the U(1) symmetry which conserves the number of particles and initialize the MPS with a product state which lies in $\mathcal{D}$. We then carefully check the relevant observables to ensure they satisfy the properties associated with $\mathcal{D}$, such as a positive height field.

By looking at various observables at stationarity, it becomes clear that there are three distinct equilibrium phases in the Fredkin model: (i) $c < 0.5$, (ii) $c = 0.5$ and (iii) $c > 0.5$.
We denote the expectation value of an observable $O$ with respect to the equilibrium state as $\braket{O}$, with
\beq
    \braket{O} = \braket{- | O | {\rm ss}} = \braket{\psi | O | \psi} . 
\eeq

The appropriate order parameter to characterize the equilibrium phases is the average area $\braket{A}$. 
In Fig.~\ref{fig: equilibrium}(a) we show $\braket{A}$ as a function of $c$ for a range of system sizes $N\in[20, 400]$. For $c < 1/2$ the area becomes minimal, while for $c > 1/2$, the area is maximal.
If we consider the area as a function of system size $N$ we find that $\braket{A}$ grows as a power law $\braket{N} \sim N^{-\beta}$, as shown in Fig.~\ref{fig: equilibrium}(b).
This reveals three distinct behaviors: the exponent $\beta$ takes the values $\beta = 1$, $3/2$ and $2$ for $c < 1/2$, $c = 1/2$ and $c > 1/2$, respectively \cite{Salberger2017}.

For each phase, we show the average of the spatial occupation profile, $\braket{n_{i}}$, and the average height field, $\braket{h_{i}}$, in Figs.~\ref{fig: equilibrium}(c,d), respectively.
For $c < 1/2$, the particles take an anti-ferromagnetic arrangement, Fig.~\ref{fig: equilibrium}(c) (red circles), thus minimizing the height and therefore the area, Fig.~\ref{fig: equilibrium}(d) (red circles). We sometimes refer to this as the {\em flat phase} (in analogy with interacting dimers \cite{Papanikolaou2007,Castelnovo2007}).

At $c = 1/2$, all configurations occur with equal probability, cf.\ \er{ss}. In terms of the RW representation of configurations this corresponds to the set of RW excursions. The average occupation, Fig.~\ref{fig: equilibrium}(c) (blue squares) interpolates between 1 and 0, 
and in the thermodynamic limit, $N\to\infty$, the average occupation density in the bulk is $1/2$ \cite{Adhikari2019}. In turn, the average height field takes a semi-circular form, Fig.~\ref{fig: equilibrium}(d) (blue squares). Note that this a phase of large fluctuations and this average height field is not representative of typical sample profiles. This is in contrast to the other two phases which are exponentially dominated by extremal area configurations, cf.\ \er{ss}. 

For $c > 1/2$, the particles (holes) localize to the left (right) edge of the system \cite{Udagawa2017}, with a sharp change in average occupation, 
Fig.~\ref{fig: equilibrium}(c) (green triangles), and with an average height profile in the shape of a tent (with a rounded top, a finite residue of the fluctuations of the $c=1/2$ phase), Fig.~\ref{fig: equilibrium}(d) (green triangles). This behavior is similar of that seen in the ASEP in an open segment with fixed boundaries \cite{De-Gier2006}. Simple arguments (see Appendix \ref{sec:profile}) give the profile \cite{Udagawa2017} 
\beq
\langle n_j \rangle = \frac{1}{\exp\left((j - N/2)/\lambda\right) + 1}
\eeq
with an inverse localization length $\lambda$,
\beq
    \lambda = \ln\left(\frac{c}{1-c}\right)^{-1}.
    \label{eq:loc}
\eeq
We sometimes refer to the $c>1/2$ phase as the {\em tilted phase} (also in analogy with interacting dimers \cite{Papanikolaou2007,Castelnovo2007}).  
Note that this shares no connection with the ``tilted generator'' later introduced in Section V.

An observable which will be of importance later is the dynamical activity $\braket{k}$, which measures the average number of configuration changes per unit time in stochastic trajectories \cite{Garrahan2007, Lecomte2007, Maes2020}. At equilibrium, it can be measured as the average escape rate, $\braket{k} = \braket{- | \mathbb{R} | \text{ss}}$, where $\mathbb{R}$ is the diagonal part of \er{W}.
We show this in Fig.~\ref{fig: equilibrium}(e)
as a function of $c$ for various system sizes $N\in[20, 400]$. It is immediately clear that the dynamical activity scales with system size (up to  small finite size effects) for $c \leq 1/2$ where occupation is spread out in equilibrium, cf.\ Fig.~\ref{fig: equilibrium}(c), leading to less constrained and therefore more dynamics throughout the entire system.
Conversely, the activity for $c>1/2$ is sub-extensive in system size as expected due to the much more inactive conditions given to the localization of the equilibrium state, cf.\ Fig.~\ref{fig: equilibrium}(c): motion is limited to the center of the lattice (the tip of the ``tent'') where particle hops are not restricted by exclusion. By fitting the activity with a linear form $\braket{k} = a + bN$ (for $c \leq 1/2$), one can extrapolate to infinite size to determine $\braket{k}/N$ in the thermodynamic limit. We show this as the black dashed line, peaking around $c\approx 0.36$.
Notice that for $c > 1/2$, the activity goes as $O(1)$ and is surpressed by the scaling in $N$.
The differences in the ``active'' ($c\leq1/2$) and ``inactive'' ($c>1/2$) dynamics are directly related to the dynamical large deviations in Sec.~IV.

\begin{figure}[t]
    \centering
    \includegraphics[width=\linewidth]{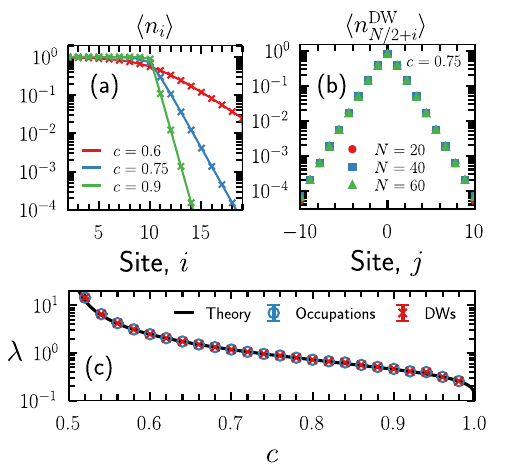}
    \caption{\textbf{Localization in the Fredkin chain.}
    (a) The occupation profile $\braket{n_{i}}$ of the steady state for $c > 0.5$ and $N = 20$. The occupations exhibit an exponential decay for $i > N / 2$.
    (b) The average domain wall occupations $\braket{n_{N/2 - i}^{\rm DW}}$ for $c = 0.75$ and $N = 20, 40, 60$. We see the same exponential decay of domain wall density as we move away from the centre of the lattice.
    (c) The localization length $\lambda$ as a function of $c$. The line shows the result from the theory, \er{eq:loc}, and the blue circles and red crosses the numerically extracted lengths from the occupation and DW profiles, respectively. The numerical data is from DMRG with $N=100$.}
    \label{fig: localization}
\end{figure}

\begin{figure*}[t]
    \centering
    \includegraphics[width=\linewidth]{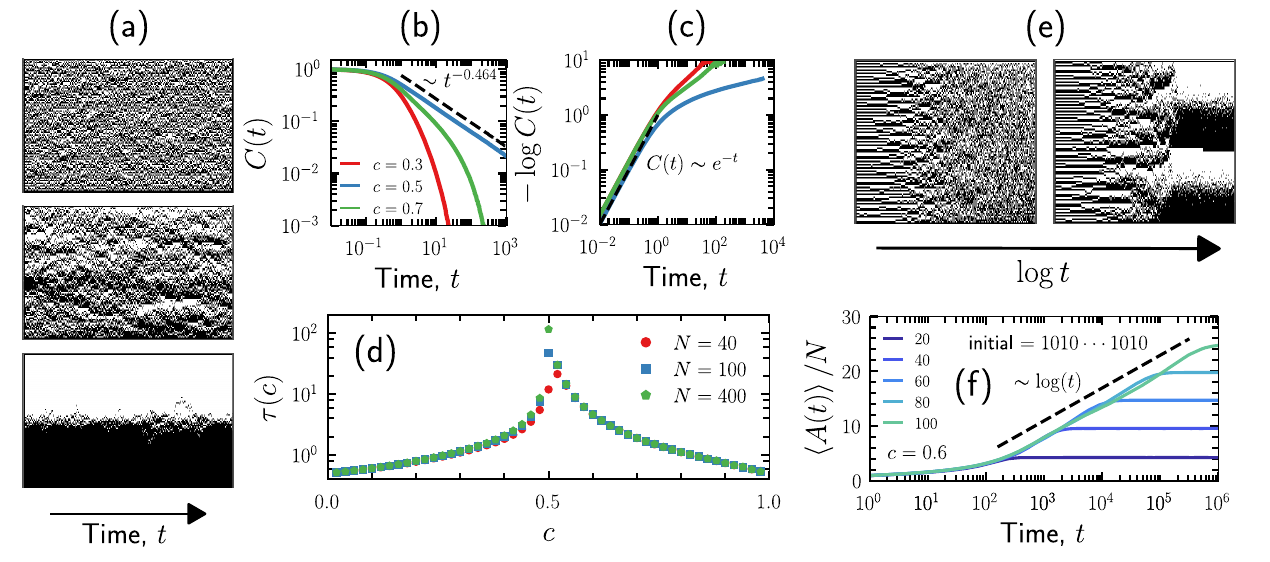}
    \caption{\textbf{Stochastic trajectories and dynamics.} 
    (a) Representative trajectories with initial states sampled from equilibrium for $c = 0.4$ (top), $c=0.5$ (middle) and, $c = 0.6$ (bottom) respectively, for system size $N = 100$ and time $t = 10^{3}$.
    (b) The autocorrelation functions \er{AC} for each of the three distinct equilibrium phases. At large times, the auto-correlator for $c = 0.5$ decays as the power law $t^{-0.464}$ (from size $N = 100$).
    (c) The same autocorrelation functions plotted on a double-log ordinate scale. For small times they show exponential decay in the three phases. For large times they take a stretched-exponential form for $c < 1/2$ and $c > 1/2$.
    (d) The numerically estimated timescales \er{timescale} as a function of $c$ (for $N = 40, 100$ and $400$).
    (e) Example trajectories after a quench from the  initial state $1010\cdots1010$ for $c = 0.5$ (left) and $c = 0.6$ (on a logarithmic time scale). The former relaxes to equilibrium quickly, whilst the latter shows hierarchical relaxation (both panels for $N = 100$ and $t=10^{5}).$
    (f) The area (scaled by system size) $\braket{A}/N$ after the same quench, for various system sizes $N\in[20, 100]$ and $c = 0.6$. 
    The dashed line shows $\log t$.
    All data is obtained using continuous-time Monte Carlo.
    }
    \label{fig: dynamics}
\end{figure*}

\subsection{Localization of the tilted phase} 
\label{secloc}

The equilibrium state for $c > 1/2$ is exponentially dominated by maximal area configurations, that is, configurations in which particles cluster towards the left edge of the system, and holes cluster at the right edge.
Figure~\ref{fig: localization}(a) shows the average occupation profile $\braket{n_{i}}$ for various $c > 1/2$: for sites beyond the halfway point, $i > N / 2$, we observe an exponential decay of the average occupation, $\braket{n_{i}} \sim e^{-i/\lambda}$. 
[Note that the same occurs for the density of holes, $1 - \braket{n_{N+1-i}}$, coming from the right, due to fact the generator \er{W} is invariant under $i \to N + 1 - i$ and $\ket{0} \leftrightarrow \ket{1}$.]

This localization can be further characterized by the density of domain walls (DWs)
\beq
    \braket{n_{i}^{\rm DW}} = \braket{n_{i}(1-n_{i+1})} + \braket{(1-n_{i})n_{i+1}}.
\eeq
This is shown in Fig.~\ref{fig: localization}(b): the DW density is close to $1$ at the centre of the lattice, and decays exponentially when moving away from it in both directions, $\braket{n_{i}^{\rm DW}} \sim e^{-|\frac{N}{2} - i| / \lambda}$. Notice that the localization is consistent for increasing system size.
As we discuss further in the next sections, exponential localization of DWs at the centre of the lattice has important implications for the dynamics in the tilted phase: particle hops can only occur when there are domain walls, and thus activity is exponentially suppressed away from to midpoint, and is sub-extensive in system size, cf.\ Fig.~\ref{fig: equilibrium}(e). 

The localization length $\lambda$ decreases with increasing $c$. We show this in Fig.~\ref{fig: localization}(c) for both particle and DW densities. The agreement with the theoretical prediction Eq.~\eqref{eq:loc} is excellent.
The numerically extracted lengths here are from DMRG with $N = 100$. For smaller system sizes, the localization length becomes comparable to system size for $c \approx 1/2$, and one might expect to see small deviations from the theoretical prediction.

\section{Typical dynamics} \label{reldyn}

\subsection{Dynamics in equilibrium}

Figure~\ref{fig: dynamics}(a) shows representative trajectories in the stationary dynamics of each of the three equilibrium phases (with the initial states sampled from equilibrium). The largest fluctuations occur for $c = 0.5$. Dynamics in equilibrium can be quantified through the (density) autocorrelation function
\beq
    C(t) = \frac{1}{N} \sum_{i=1}^{N} \frac{\braket{n_{i}(0)n_{i}(t)} - \braket{n_{i}}^{2}}{\braket{n_{i}} - \braket{n_{i}}^{2}} ,
    \label{AC}
\eeq
which provides a measure of the memory of a initial configuration after time $t$ in an equilibrium trajectory.
We show $C(t)$ for the three equilibrium phases in Figs.~\ref{fig: dynamics}(b,c). It is apparent from 
Fig.~\ref{fig: dynamics}(b) that for $c = 1/2$ the autocorrelation decays asymptotically as a power law, with a numerically extracted exponent of just under a half. 
This power law decay can also be observed for intermediate times at $c > 1/2$ [corresponding to fluctuations of the top of the ``tent'', cf.\ Fig.~\ref{fig: equilibrium}(d)], with this intermediate regime becoming longer as $c$ gets closer to $1/2$.
While at short times decay is exponential, see 
Fig.~\ref{fig: dynamics}(c), for long times relaxation is stretched exponential in both the flat and tilted phases. These are signatures of slow dynamics. 

We can extract a timescale for relaxation of correlations from $C(t)$ from its integral,
\beq
    \tau_{\rm eq} = 
        \int_{0}^{\infty} 
            C(t) dt .
    \label{timescale}
\eeq
This is shown in Fig.~\ref{fig: dynamics}(d) for a range of $c$. This equilibrium timescale spikes at $c = 1/2$, as expected from the slow law decay of $C(t)$.
Notice that the spike is less sharp for smaller system sizes due to the finite size effects from the boundaries.

\subsection{Relaxation towards equilibrium}

Also of significance is the relaxation towards the equilibrium state when starting from non-equilibrium  conditions. We explore this behavior by considering dynamics following from an initial state of minimal area, $x_0 = 1010\cdots1010$, corresponding to a quench from deep in the flat phase ($c_0 \gtrsim 0$) to finite $c > 0$. When $c < 1/2$, equilibrium is achieved quickly as the initial state is not far from typical states in the flat phase. Interesting non-equilibrium dynamics occurs when quenching to $c=1/2$ or to the tilted phase at $c>1/2$. In Fig.~\ref{fig: dynamics}(e) we show two relaxation trajectories, one for $c = 1/2$ (left) and another for $c = 0.6$ (right)
\footnote{The lack of visible fluctuations for $c=0.5$ is a consequence of the logarithmic scale in time.}. The system size is $N = 100$ and the overall time of trajectories $t = 10^{5}$ (note that the time axis is shown on a logarithmic scale). For the case of a quench to $c = 1/2$, after a slow early regime, equilibrium is reached in reasonable times.

For a quench to $c > 1/2$, we observe a slow hierarchical relaxation, with a progressive coarsening of clusters of particles and holes. The target state is a tilted one, cf.\ Figs.~\ref{fig: equilibrium}(d), and in the height representation this hierarchical process is the merging of smaller tents in the profile into larger ones. Due to the constraint, \er{fi}, local configurations of $\cdots0011\cdots$, corresponding to troughs in the height field, 
are locally trapped, and require particles from the right edge of clusters to diffuse away to allow clusters to merge. This process is exponentially scarce in the separation distance, as occupations are exponentially localized, cf.\ Sec.~\ref{secloc}.

The time to complete each stage of relaxation in the tilted phases grows exponentially with the stage. This hierarchy is evident in the growth of the average area normalized by system size, $\braket{A(t)}/N$, as shown in Fig.~\ref{fig: dynamics}(f), where we see the area increasing logarithmically in time. This reveals the hierarchical nature of the relaxation process: while smaller systems may have relaxed to equilibrium, larger systems require the merging of larger clusters, and so the growth of the area continues.

\begin{figure*}[t]
    \centering
    \includegraphics[width=\linewidth]{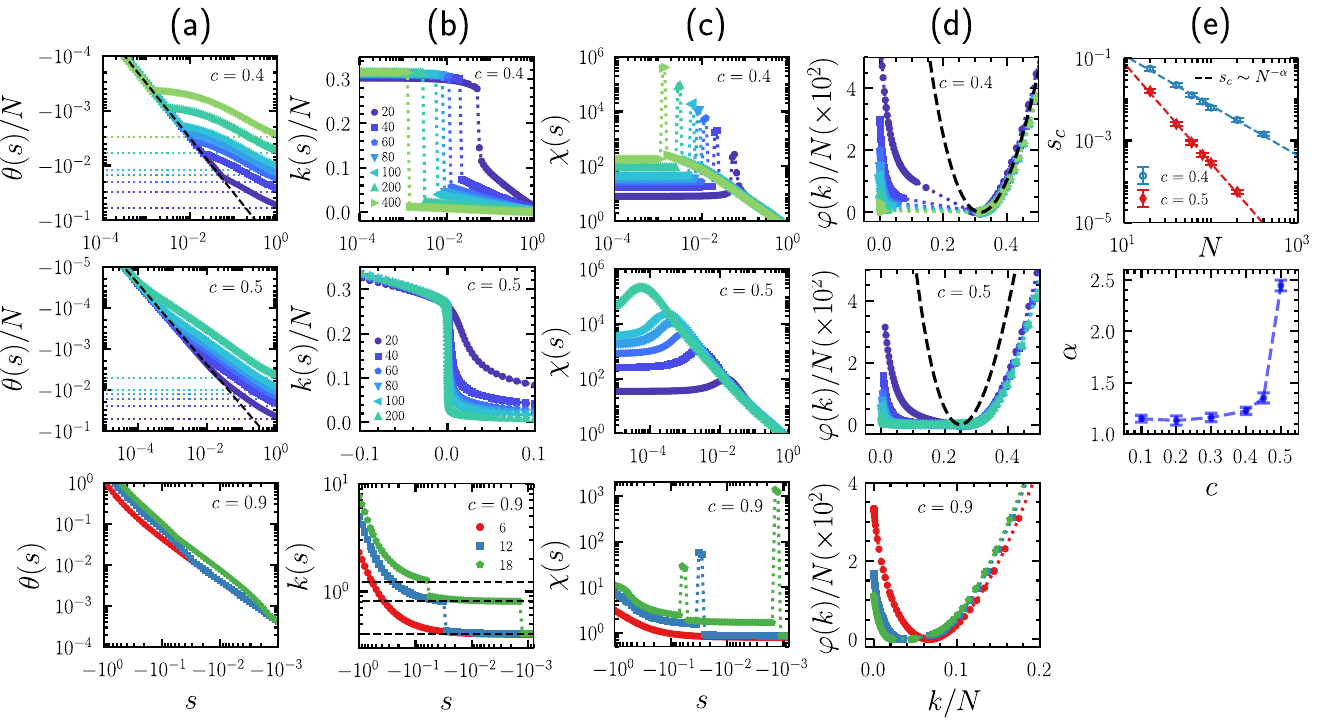}
    \caption{\textbf{Finite size scaling of dynamical LD transitions.}
    The dynamical LD statistics for each equilibrium phase. The top (middle) row of (a-d) shows $c=0.4$ ($c=0.5$) with $N\in[20, 400]$ obtained via DMRG and the bottom row shows $c=0.9$ obtained through ED.
    (a) The SCGF $\theta(s)$ as a function of $s$. The upper and middle panels are scaled by system size, with dotted lines showing the value for $s\to\infty$ and the dashed line showing the linear response prediction (see the main text).
    (b) The average dynamic activity  $k(s)$ as a function of $s$. The top and bottom panels are shown on log-log scales, whereas the middle one is shown in linear scale. The dashed lines in the bottom panel correspond to integer multiples of $k(0)$.
    (c) The dynamical susceptibility $\chi(s) = \theta''(s)$ as a function of $s$.
    (d) The LD rate function scaled by system size $\varphi(k)/N$ as a function of activity $k$. The black dashed lines show a Poisson distribution with mean $k(0)/N$ in the thermodynamic limit $N\to\infty$, extrapolated from finite-size DMRG data. 
    (e) We estimate the critical point as a function of system size from the peaks of the dynamical susceptibility for $c = 0.4, 0.5$ in the top panel. The dashed lines shows a fitted power law $s_{c} \sim N^{-\alpha}$, with the bottom panel showing the obtained $\alpha$ for various $c$.
    }
    \label{fig: LD_statistics}
\end{figure*}

\section{Dynamical large deviations} \label{dynLDs}

We now study the statistical properties of the stochastic trajectories $\omega_{t} = x_{0:t}$ of the Fredkin model, in particular the LD statistics of dynamical observables. 

If $K(\omega_{t})$ is a trajectory observable, the probability of it taking a value $K$ is
\beq
    P_{t}(K) = \sum_{\omega_{t}} \pi(\omega_{t})\delta[K(\omega_{t}) - K],
\eeq
where $\pi(\omega_{t})$ is the probability of the trajectory $\omega_{t}$ being realized under the stochastic dynamics. For a dynamical observable $K$ that is time-extensive, in the long-time limit, the probability of $K$ obeys a LD principle   \cite{Touchette2009, Garrahan2018, Jack2020, Limmer2021}
\begin{align}
    P_{t}(K) \asymp e^{-t\varphi(K/t)}
\end{align}
where the function $\varphi(k)$ is called the LD rate function. The above asymptotic equality holds as long as the spectral gap is non-vanishing (which it is in the Fredkin model for finite size $N$ \cite{Movassagh2016}). A LD principle also holds for the moment generating function (MGF)
\begin{align}
    Z_{t}(s) 
        = \sum_{K} P_{t}(K) \, e^{-sK} 
        = \sum_{\omega_{t}} 
            \pi(\omega_{t}) \, e^{-sK(\omega_{t})}
        \asymp e^{t\theta(s)} 
        \nonumber
\end{align}
where $\theta(s)$ is the scaled cumulant generating function (SCGF) whose derivatives at $s=0$ give the cumulants of $K$, scaled by time \cite{Touchette2009}.
In analogy with what occurs in equilibrium thermodynamics, the rate function and SCGF are related by a Legendre transform 
\begin{align}
    \theta(s) = -\min_{k}[sk+\varphi(k)] 
\end{align}

We consider as observable $K$ the dynamical activity. Its SCGF is given by largest eigenvalue of the {\em tilted generator}, $\W_{s}$, which for the Fredkin model reads
\begin{align}
    \W_s = \sum_{i = 2}^{N-2} 
        &
        f_{i} 
        \left\{    
            c 
            \left[ e^{-s}
                 \sigma^{+}_{i} \sigma^{-}_{i+1} 
                 - (1-n_{i}) n_{i+1} 
            \right]
        \right.
    \label{Ws}
    \\ 
        & 
        \phantom{f_{i}}
        \left.
            + (1-c) 
            \left[ e^{-s}
                \sigma^{-}_{i} \sigma^{+}_{i+1} 
                - n_{i} (1-n_{i+1}) 
            \right] 
        \right\} ,
        \nonumber
    \end{align}
with $s$ being {\em counting field}. As $\W_s$ is in general non-Hermitian, the leading eigenvalue $\theta(s)$ has right and left eigenvectors $\ket{r_{s}}$ and $\bra{l_{s}}$. 

We can write the generator in a Hermitian form with the same similarity transformation as before, \er{sim}, 
\begin{align}
    \H_s &= 
        - \sum_{i=2}^{N-2} 
            f_{i} 
            \left[ e^{-s}
                \sqrt{c(1-c)}
                \left(
                    \sigma^{+}_{i} \sigma^{-}_{i+1} + \sigma^{-}_{i} \sigma^{+}_{i+1}
                \right)
            \right.
    \label{Hs}
    \\
            & 
            \left.    
                \phantom{\sqrt{c(1-c)}}
                - c (1 - n_{i}) n_{i+1}
                - (1 - c) n_{i} (1 - n_{i+1}) 
            \right] ,
    \nonumber
\end{align}
with ground state $\H_{s}\ket{\psi_{s}} = -\theta(s)\ket{\psi_{s}}$, related to the leading eigenvectors of $\W_s$ by 
\beq
    \ket{\psi_{s}} = 
        \sum_{x} \sqrt{l_{s}(x)r_{s}(x)} \ket{x},
    \label{psi}
\eeq
where $l_{s}(x) = \braket{l_{s} | x}$ and $r_{s}(x) = \braket{x | r_{s}}$.

\begin{figure*}[t]
    \centering
    \includegraphics[width=\linewidth]{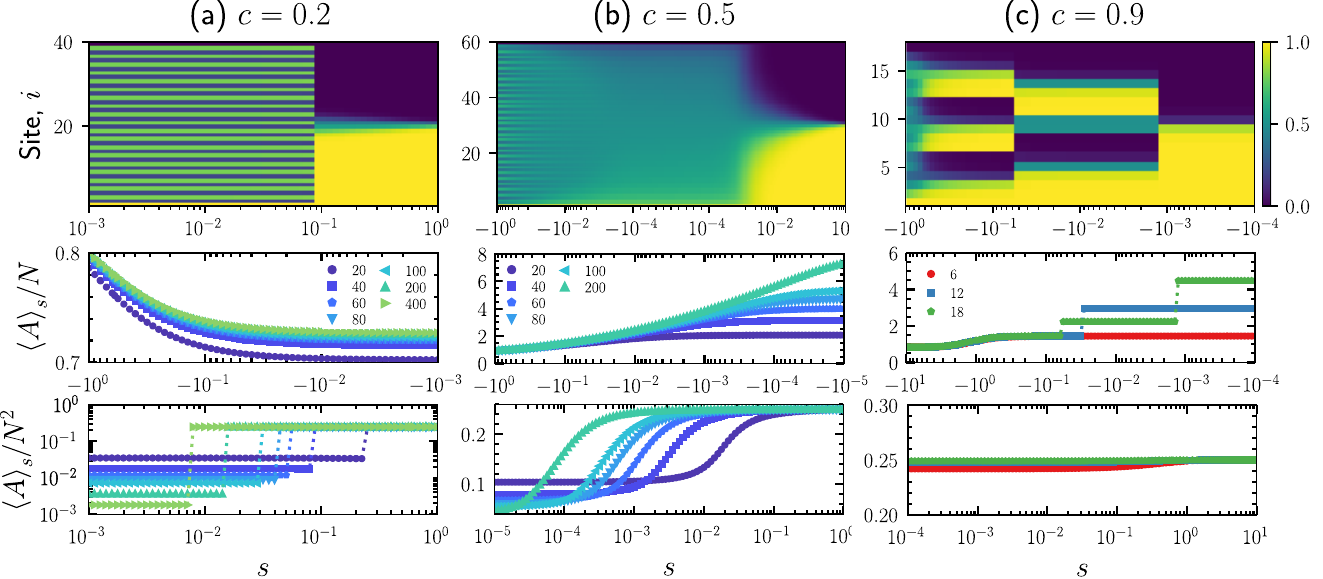}
    \caption{\textbf{Structural properties of the LDs.}
    We show observables for each equilibrium phase with (a) $c = 0.2$, (b) $c = 0.5$ and (c) $c = 0.9$.
    The top row shows the average occupations $\braket{n_{i}}_{s}$ for site $i$ (with differing system sizes and ranges of $s$).
    The middle row shows the area scaled by system size $\braket{A}_{s} / N$ for $s < 0$.
    Finally, we show the area scaled by system size squared $\braket{A}_{s} / N^{2}$ for $s > 0$ in the bottom row.
    }
    \label{fig: LD_structure}
\end{figure*}

\subsection{Active-inactive trajectory transitions at $c \leq 1/2$}

From the ground state of \er{Hs} we can study statistical properties of the trajectory ensemble of the Fredkin model for long-time trajectories. We do this by  means of numerical tensor networks along the lines of similar recent work in KCMs
\cite{Banuls2019, Helms2019, Helms2020, Causer2020, Causer2021, Causer2021b}.
Figure~\ref{fig: LD_statistics} shows the LD statistics obtained numerically. 
The top row gives this for the flat phase at $c = 0.4$, and the middle row for the $c = 1/2$ phase. These results are for system sizes in the range $N\in[20, 400]$ obtained using DMRG. 

Column (a) of Fig.~\ref{fig: LD_statistics} shows the SCGF as a function of $s=0$ for a range of sizes.  
For small $s \gtrsim 0$, the SCGF follows linear response (LR), $\theta(s) \approx - s k(0)$, where $k(s) = -\theta'(s)$ is the average dynamical activity in the ensemble tilted by $s$, shown in column (b). The LR prediction is shown by the dashed black line for $N\to\infty$,  calculated by fitting the dynamical activity for finite sizes at $s = 0$ with a power law and extrapolating. 
Notice that at some $s_{c}(N)>0$, which becomes smaller for increasing $N$, the behavior deviates from LR to one which no longer scales with $N$ (this is most apparent for $c < 1/2$). The step in the average activity, Fig.~\ref{fig: LD_statistics}(b, top and centre), indicates a phase transition between dynamical phases of high and low activity. The change in activity tends to a discontinuity with increasing size,  indicative of a first-order transition.

The point $s_{c}(N)$ at which the crossover occurs at finite size can be estimated from the peak in the corresponding dynamical susceptibility, $\chi(s) = \theta''(s)$,  shown in column (c) of Fig.~\ref{fig: LD_statistics}. As the system size is increased, the crossover point shifts towards $s = 0$ and becomes sharper. 
The change in dynamics can be seen in the broadening of the LD rate function $\varphi(k)$ around the equilibrium average, shown in column (d) of Fig.~\ref{fig: LD_statistics}. The rate functions show the characteristic Maxwell construct of a first-order transition between two phases, and active one with large $k$ and an inactive one with vanishing $k$.
Note that while the transition in activity looks less sharp for $c=0.5$, we expect to recover the usual first order behaviour for increasing system sizes as seen by the broadening of the rate function.

For $c \leq 1/2$, the location of the crossover can be fit by a power law $s_{c}(N) \sim N^{-\alpha}$. The upper panel of column (e) in Fig.~\ref{fig: LD_statistics} shows this for $c = 0.4$ and $c = 1/2$. The lower panel of column (e) shows 
the dynamical exponents $\alpha$ as a function of $c$. When $c$ is far from $1/2$, we have approximately $\alpha \approx 1.2$. The exponent increases quickly as we approach $c = 1/2$, to around $\alpha \approx 2.5$, a value similar to that found in other exclusion processes \cite{Causer2020}. It could be that for values close to (but not equal to) $c = 1/2$, the measured exponent would be lower if we accounted for larger system sizes.

\subsection{Dynamical phases for $c>1/2$}

Obtaining accurate estimates of $\theta(s)$ for $c > 1/2$ at large system sizes is difficult due to a proliferation of dynamical phases. In particular, it is hard for DMRG to converge to the correct phase due to a large density of states. For this reason, for $c > 1/2$ we limit our studies to system sizes $N = 6, 12, 18$ with large $c = 0.9 \gg 1/2$ which allows us to effectively study the hierarchy of dynamical phases using exact diagonalization (ED)
\footnote{
    We construct the basis so that only the dynamically connected configurations within the sector are included.
    }.
The bottom row of Fig.~\ref{fig: LD_statistics} shows these results. 

Since the typical dynamics ($s=0$) of the tilted $c>1/2$ phase is itself inactive, cf.\ Fig.~\ref{fig: equilibrium}(e), we expect transitions to the active phase to occur at $s < 0$ for finite size systems. In fact, column (b) of Fig.~\ref{fig: LD_statistics} shows several points where the behavior of the SCGF changes. The number of these points seems to increase with system size.
In each case, this change in behavior corresponds to transitions in the dynamics. At each of these points we see a sharp drop in the activity, this becoming sharper with increasing $N$. The values at which the activity plateaus 
are multiple integers of the equilibrium activity, $k(s=0)$, and are shown by the black dashed lines. With the limited range of sizes accessible via ED it is not possible to do a finite-size scaling analysis as we did for $c<1/2$. From the systems studied we observe that the first away from equilibrium inactive behavior happens at increasing $s$ (that is, getting closer to $0$) for increasing $N$, which shows in the flattening of the rate function, see bottom panel in Fig.~\ref{fig: LD_statistics}(d). 

\begin{figure}[t]
    \centering
    \includegraphics[width=\linewidth]{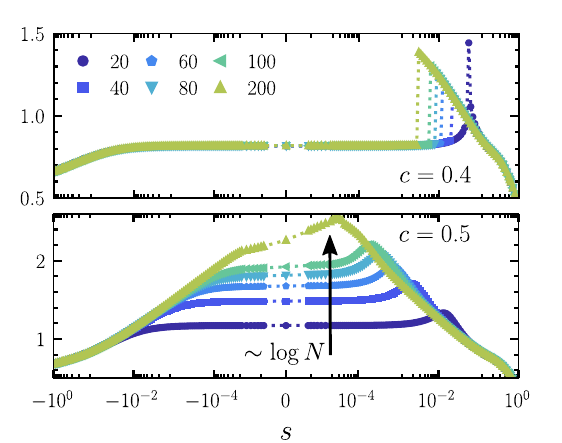}
    \caption{
        \textbf{Entanglement entropy of the LD eigenvectors.}
        The bipartite entanglement entropy $S_{E}(s)$ for $c = 0.4$ (top) and $c = 0.5$ (bottom) and system sizes $N\in[20, 200]$.
        For $c = 0.5$, the entanglement entropy scales as approximately $S_{E} \sim \log N$.
        }
    \label{fig: entanglement}
\end{figure}

\subsection{Structural properties of the dynamical large deviations}

The difference in the behavior of the various dynamical phases also manifests in the structural properties of the configurations  visited by the trajectories. The eigenvector $\ket{\psi_{s}}$ obtained from either DMRG and ED contains the probability amplitudes for each configuration, making it easy to calculate averages of configuration observables $\O(x)$ in the tilted ensemble
\footnote{
    The meaning of this is the following: at a given value of $s$ what we are exploring is the ensemble of trajectories $\omega_t$ with probabilities reweighed from those of the original dynamics as $\pi_s(\omega_t) = Z_t^{-1}(s) \pi(\omega_t) e^{-s K(\omega_t)}$. 
    For long time trajectories, $t \gg 1$, the probability of finding a configuration $x_{t'}$ at an intermediate time $1 \ll t' \ll t$ (i.e., far from both initial and final time boundaries) is given by $l_s(x) r_s(x) = \psi_s(x)^2$. Expectation values such as $\braket{A}_s$ provide information about the properties of configurations at the ``mid point'' of long-time atypical trajectories in the ensemble characterized by $s$. 
},
\begin{align} 
    \braket{\O}_s 
        = \bra{L_s} \O \ket{R_s} 
        = \sum_x \O(x) \psi_s(x)^2
\end{align} 

In Fig.~\ref{fig: LD_structure} we show the local occupations $\braket{n_{i}}_{s}$ (top panels), and the average area $\braket{A}_{s}$ (middle and bottom panels), for (a) $c = 0.2$, (b) $c = 1/2$ and (c) $c = 0.9$.
It is clear that the limit of large activity ($s < 0$ with $|s|$ large) particles spread out in order to maximize the activity. This is evident by the average area $\braket{A}_{s}$, which scales linearly with system size $N$, resembling the structures associated with the equilibrium flat phase for $c < 1/2$. Thus, the active phase for all values of $c$ is also a structurally flat one. Conversely, in the inactive limit for all values of $c$ (large $s > 0$) particles cluster at the left edge of the system and maximize the area,  which scales as $N^{2}$. Thus, irrespective of the equilibrium static phase, the inactive phase of the dynamics is structurally ``tilted''. 

Interestingly, we observe very sharp transitions for $c \neq 1/2$ even at smaller sizes - this is unusual when compared to other constrained models \cite{Banuls2019, Causer2020}. This could be due to the sharp transition in activity at equilibrium, cf.\ Fig.~\ref{fig: equilibrium}.
Indeed, for $c > 1/2$ we notice sharp structural changes at the location of the sharp points of 
Fig.~\ref{fig: LD_statistics}. It is clear that the corresponding structures are related to the assembly of excited states at equilibrium ($s = 0$) obtained by joining multiple ground states of smaller system sizes (compare to what occurs in the excited states of the quantum East model \cite{Pancotti2020}). Of course this makes sense, as despite the scarcity of the configurations associated with these states, they have large lifetimes (as discussed in Sec.~III) with impactful consequences on the relaxation behavior.

\begin{figure}[t]
    \centering
    \includegraphics[width=\linewidth]{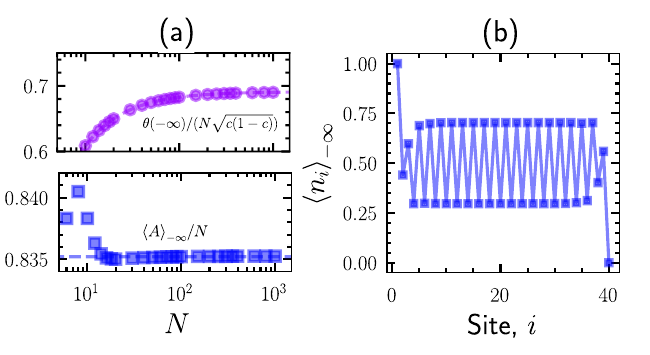}
    \caption{
        \textbf{Extreme active limit.}
        (a) The rescaled SCGF $\Tilde{\theta}/N$ (top) and the area $\braket{A}_{-\infty} / N$ (bottom) as functions of $N$ measured via DMRG. We fit the SCGF as $a + bN^{-1}$ allowing us to extrapolate the value in the thermodynamic limit $N\to\infty$ (see main text), shown by the dashed line.
        The value for the area quickly settles with increasing system size, indicated by the dashed line.
        (b) The occupation profiles $\braket{n_{i}}_{-\infty}$ for $N=40$.
        }
    \label{fig: limits}
\end{figure}

\subsection{Entanglement entropy}
We now consider the bipartite von Neumann entanglement entropy of the MPS approximations to \er{Hs}.
We partition the system into two subsystems A and B, which denote the spins $i\in [1, N/2]$ and $i\in [N/2+1, N]$ respectively. The bipartite entanglement entropy between the two partitions is then calculated by
\beq
    S_{E}(s) = -\Tr[\rho_{\rm A}\log\rho_{\rm A}],
\eeq
where $\rho_{A} = \Tr_{B}[\rho]$ denotes the reduced density matrix for A, and $\rho = \ket{\psi_s}\bra{\psi_s}$ is the density matrix for the full system.

The Hamiltonian \er{Hs} exhibits a ground state phase transition in the bipartite entanglement entropy for $s = 0$.
In particular, it scales as $S_{E}(0) \sim \log N$ for $c = 1/2$ and $S_{E}(0) \sim 1$ for $c \neq 1/2$ \cite{Movassagh2016, Salberger2016, Sugino2018, DellAnna2019a}.
We now extend this analysis to $s\neq 0$.
Figure~\ref{fig: entanglement} shows the entanglement entropy for increasing system size $N\in [20, 200]$ for $c = 0.4$ (top) and $c = 0.5$ (bottom) and a range of $s$. 
Notice that for $c < 1/2$, the entropy obeys an area law for all $s$ (although we observe spikes around the transition from active to inactive dynamics).
For $c = 1/2$, we observe for large magnitude $s < 0$ the states clearly also obey an area law. As $s$ approaches $s=0$, the entanglement entropy appears to grow significantly towards $S_{E}(0)$, and looks to scale logarithmically.
It is important to note however we only show a small range of system sizes, and it is most likely that for some fixed $s < 0$, the entanglement entropy will be bounded as $N\to\infty$, and thus obeys an area law. This can be seen by the ``branching'' behaviour seen in Fig.~\ref{fig: entanglement}. 
An important consequence is that for large enough $N$, one is able to construct a state with arbitrarily high entropy by tuning the value of $s$ towards $s=0$.
For the inactive phase $s > 0$, $\psi_{s}$ clearly also obey an area law, again with the entanglement entropy spiking as $s$ approach $s = 0$.

\subsection{Limits of maximal and minimal activity}

The limit of maximum activity is that at $s \to -\infty$.
In this limit, the diagonal parts of $\mathbb{W}_{s}$ (and $\mathbb{H}_{s}$) are suppressed and only the off-diagonals are left. Notice that for $\mathbb{H}_{s}$, the dependence on $c$ falls out as a prefactor.
As the tilting in $\W_s$ grows exponentially with $-s$ for negative $s$, we rescale the SCGF as 
\beq
\tilde{\theta} 
    = \lim_{s\to-\infty}
        \frac{e^{s}\theta(s)}{\sqrt{c(1-c)}},
\eeq
when taking the limit. The (rescaled) eigenvalue $\tilde{\theta}$ coincides with the (similarly rescaled) dynamical activity. We show this in Fig.~\ref{fig: limits}(a) as a function of $N\in[10, 400]$ (circles, shown divided by $N$), and fit it with 
the function of $a N + b$ (blue dashed line, shown divided by $N$). By extrapolating to infinity, we find that 
\beq
    \lim_{N\to\infty}\Tilde{\theta} / N \approx 0.691 .
\eeq
The average area $\braket{A}_{-\infty}$, see  Fig.~\ref{fig: limits}(a) takes an almost constant value, with small fluctuations for small system sizes,
\beq
    \lim_{N\to\infty}\braket{A}_{s}/N \approx 0.835.
\eeq
Notice that the area scales linearly with system size, and is similar to the equilibrium states found for $c<1/2$.
This is further seen from the average occupations $\braket{n_{i}}_{-\infty}$, see Fig.~\ref{fig: limits}(b), showing the antiferromagnetic pattern of the flat equilibrium phase. 

The opposite limit of $s\to\infty$ gives the most inactive state. 
In this limit only the diagonal escape rate part of \er{Ws} (or \er{Hs}) remains and each configuration $x\in\D$ is an eigenstate. The configurations with the smallest escape rates dominate. Depending on $c$ and $N$, this is either the maximal area (i.e., fully tilted) configuration, $1111\cdots0000$, which has escape rate $R=2(1-c)$, or the minimal area configuration $1010\cdots 1010$, which has escape rate $R = c(N-2)$. The latter dominates if $N > 2c^{-1}$, and the former dominates if  
$N < 2c^{-1}$ (with degeneracy at $N = 2c^{-1}$).

\section{Conclusions} \label{conc}

Here we have provided a detailed study of the statics and dynamics of the stochastic Fredkin model. Despite being one-dimensional and having local transition rules, this model displays phase transitions between three distinct equilibrium phases. This is a consequence of the constraints in the dynamics which restrict the state space to that of random walk excursions, with these static transitions controlled by the asymmetry in the particle hopping rates. Two of these phases are ordered, one being flat and another one tilted (in terms of the height field representation), with an intermediate disordered and fluctuating phase. This phase behavior is in some ways reminiscent of interacting two-dimensional dimer coverings \cite{Papanikolaou2007, Castelnovo2007, Stannard2012}. 

The constraints in the local transitions of the Fredkin model lead to a rich dynamics, both in equilibrium and in the relaxation after a quench. This richness can be seen as a consequence of a non-trivial phase structure of the ensemble of stochastic trajectories. Using numerical matrix product states with DMRG, we compute the large deviations of the dynamical activity and show the existence of active-inactive space-time phase transitions, something that is also observed in other KCMs. The overall picture is one where the 
static phases extend into dynamical ones, with the 
flat phase being also a dynamical active phase, and the tilted phase a dynamical inactive one, with first order transitions between them. 

There are many possible continuations of the work here. One is to go beyond one-dimension. As an initial step, in Appendix \ref{tilings} we propose a two-dimensional generalization of the Fredkin model: by focusing on the fact that Fredkin configurations are random walk excursions, we proposed a two-dimensional model in terms of packed dimers on the honeycomb lattice with constraints in the the dynamics which enforce configurations to be ``excursion surfaces''. It will be interesting to study this and similar stochastic models in future work. Another interesting area of exploration would be to study the Hamiltonian \er{Hs} under unitary dynamics, in analogy with recent work that studied other quantum KCMs. 
As occurs with the quantum East model \cite{Pancotti2020}, 
we expect the constraints in Fredkin models to provide mechanisms for localization and non-thermal eigenstates. We hope to report on this in the near future.

\begin{acknowledgments}
    We acknowledge financial support from EPSRC Grants EP/R04421X/1 and EP/P034616/1 and the Leverhulme Trust Grant RPG-2018-181. 
    We acknowledge access to the University of Nottingham Augusta HPC service. 
\end{acknowledgments}

\appendix

\begin{figure*}[t]
    \centering
    \includegraphics[width=\textwidth]{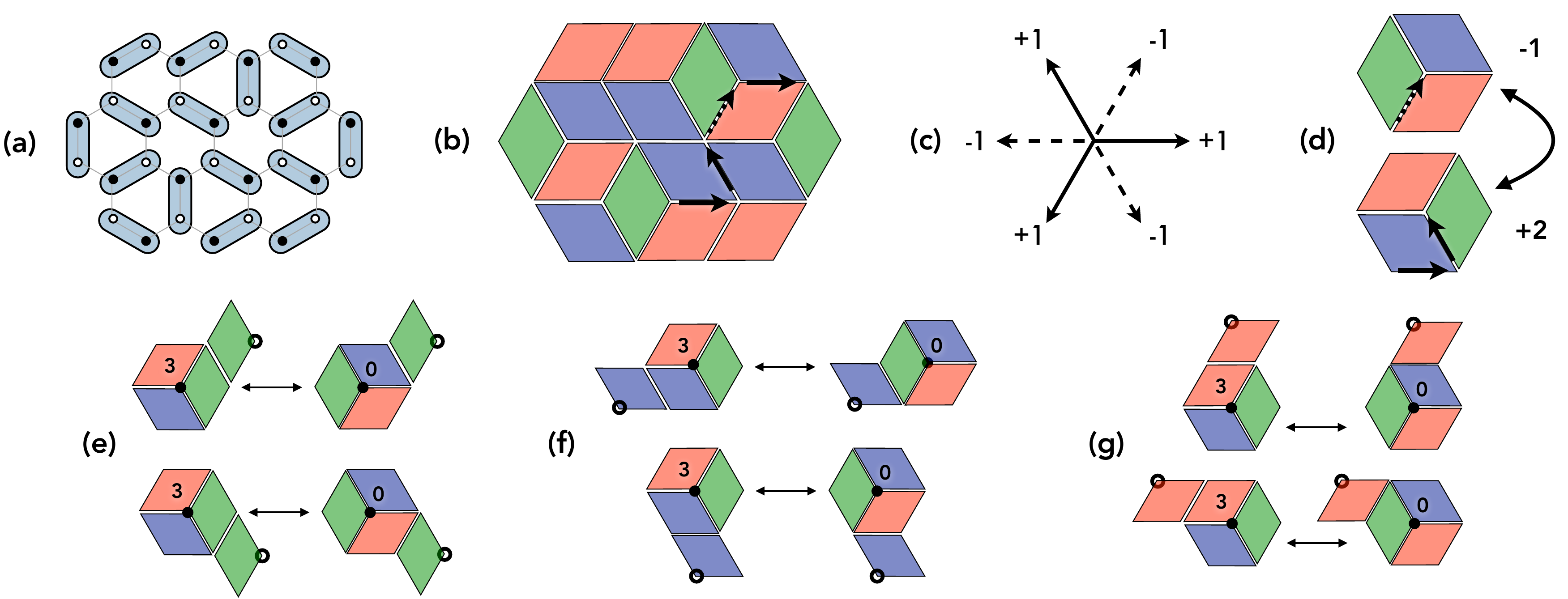}
    \caption{
        \textbf{Two-dimensional generalization of the Fredkin model.}
        (a) Dimer covering of the honeycomb lattice. (b) Equivalent rhombus tiling of the plane. (c) Definition of the height field: given a tiling, moving along the edges of the rhombi the height increases of decreases by one unit as shown. For example, the path in (b) shows that the start and end points have a height difference of $+2$. (d) The elementary local moves that preserve the ``perfect tiling'' character (i.e., no tiling defects, or no monomers in the dimer representation) are rotations of a triplet of tiles forming an elementary hexagon. These are the dimer/tiler equivalent of the particle-hole exchange in the SEP. These moves change the height field of the central site by three units.
        (e-g) Constrained moves: requiring the presence of the extra tile guarantees that the height of the central site (indicated by the filled circle) never goes below the lowest height of the arrangement (indicated by the open circle). These are the two-dimensional equivalents of the allowed moves in the Fredkin chain, see Fig.~\ref{fig: representations}(a). 
        }
    \label{fig: tilings}
\end{figure*}

\section{The density profile for $c>1/2$}
\label{sec:profile}

Simple statistical mechanical considerations can be used to compute the density profile for $c>1/2$ in the thermodynamic limit. Recall that the probability of a configuration $x$ is weighted by a factor that depends on the area $A(x)$ under the path
$$
P(x)\propto \left(\frac{c}{1-c}\right)^{\frac{1}{2}A(x)} = \exp\left(-\beta A(x)\right)
$$
where $\beta \equiv \frac{1}{2}\log [(1-c)/c]$. The entropy associated with a configuration is just the sum of the binary entropies
$$
S(x) = -\sum_i \left[n_i\log n_i + (1-n_i)\log(1-n_i)\right].
$$
After writing the area as
\beq
    A(x) = \sum_{i=j}^{N} h_{j}(x) = \sum_{i=j}^{N} (N+1-j) \, Z_j
\eeq
($Z_i=2n_i - 1$) we arrive at the `free energy' 
$$
\sum_{j=1}^N\left[ (\xi- j)Z_j  \right] - \beta^{-1}S
$$
where $\xi$ is a Lagrange multiplier introduced to fix the overall particle number $\sum_j n_j$. Extremizing the free energy gives
$$
Z_j = \tanh\left(\beta\left[\xi - j\right]\right).
$$
Thus $Z_j$ has a domain wall profile with a location $\xi$ that is determined by the particle number ($\xi=N/2$ for half filling), and a width
$$
\lambda\equiv (2\beta)^{-1} = 
\left[
    \ln{\left(\frac{c}{1-c}\right)}    
\right]^{-1} .
$$

\section{Possible two-dimensional generalization}
\label{tilings}

The height representation of the Fredkin model suggest several possible generalizations to two dimensions by analogy with dimer coverings. One possibility is the following.

Consider a fully packed dimer covering of the honeycomb lattice, see Fig.~\ref{fig: tilings}(a), where each link connecting any two neighboring sites of the lattice is occupied by a dimer. Such coverings have a height representation in terms of a height field $h_{i,j}$, which becomes apparent in the equivalent rhombus tiling of the plane, see Fig.~\ref{fig: tilings}(b): from some origin $(0,0)$ where $h_{0,0} = 0$, the height of a site is computed by moving along the edges of the rhombi with $\Delta h$ at each step according to Fig.~\ref{fig: tilings}(c). For example, in the covering of Fig.~\ref{fig: tilings}(b) the two initial and final sites connected by the path with the arrows differ in height by $\Delta h = 2$. For fully packed dimer configurations (also called ``perfect tilings'') it is easy to verify that any path that connects two sites gives the same height difference and the height field is uniquely defined. Honeycomb dimer coverings (rhombus tilings) therefore define surfaces in two dimensions. In a configuration with an equal amount of the three kind of tiles the height field is pinned at zero at the boundaries (for example in three sites at angles of $2 \pi/3$ within a hexagonal region). This is a two-dimensional version of the one-dimensional height field from a lattice of particles and holes at half filling which is bound to return to the origin. 

In the one-dimensional case the elementary local move that preserves the filling fraction is to exchange a particle with an adjacent hole. The analogous move for a rhombus tiling is shown in Fig.~\ref{fig: tilings}(d) and corresponds to rotating a triplet of tiles forming an elementary hexagon. This move changes the height of the central site by $\Delta h = \pm 3$. In order to prevent the height field from becoming negative, which is the defining property of the dynamics of the Fredkin model, transitions like those of Fig.~\ref{fig: tilings}(d) have to be constrained, cf.\ Fig.~\ref{fig: representations}(a). Figures~\ref{fig: tilings}(e-g) show the corresponding allowed transition in the two-dimensional case: the exchange of tiles is only possible if either of the extra green/blue/red tile as in arrangement (e/f/g), respectively, is present, and not allowed otherwise. This constraint implies that in the transition the height of the site at the centre of the hexagon cannot go below that of the site indicated by a circle. With this dynamical rules it is guaranteed that the height field of the dimer/rhombus arrangement never becomes negative at any point, a two dimensional version of the RW excursions that define the configurations of the Fredkin model. Furthermore, giving different rates to the forward and backward moves in Figs.~\ref{fig: tilings}(e-g) should lead to flat and tilted phases weighted by the volume under the surface.

\bibliographystyle{apsrev4-1}

\end{document}